\documentclass[12pt]{article}
\usepackage{amsxtra,amssymb,amsthm,amsmath,latexsym}

\theoremstyle{plain}

\def\oH{\buildrel\circ\over H}
\def\oH1{\buildrel\circ\over H\kern-.02in{}^1}

\begin{document}


\title{ A counterexample to the uniqueness result of Cox and Thompson
}


\author{ A.G. Ramm \thanks{Math subject classification: 34R30; PACS:  
03.80.+r. 03.65.Nk } 
\thanks{This paper was written when the author
was visiting Institute for Theoretical Physics, University of
Giessen. The author thanks DAAD for support and Professor W.Scheid for
discussions} \\
 Mathematics Department, Kansas State
University, \\
Manhattan, KS 66506-2602, USA\\
ramm@math.ksu.edu\\
}

\date{}

\maketitle\thispagestyle{empty}

\begin{abstract} A counterexample is given to the
uniqueness result given in the paper by J.Cox
and K.Thompson, 
Note on the uniqueness of the solution of an equation of interest
in inverse scattering problem,  J. Math.Phys., 11, N3, (1970),
815-817.

\end{abstract}


\section{Introduction} 

In \cite{CT} the authors claimed that the integral equation
$$h(s) =-\int_0^r g(s,t)h(t)t^{-2}dt,  
   \eqno{(1.1)}$$
where $h(0)=0$, has only the trivial solution for all $r>0$. Here
$$g(s,t)= \sum_{m\in S}\gamma_m u_m(t)v_m(s),\quad s>t,$$ 
the set
$S$ is a finite set of distinct real numbers from the
interval $(-0.5, \infty)$, $u_m$ and $v_m$ are the regular and
irregular Bessel-Riccati functions, defined
e.g. in \cite{CS},  $g(s,t)$ is symmetric, and $\gamma_m$ satisfy
the following equation: 
$$\sum_{m\in S}\gamma_m [ m(m+1)-l(l+1)]^{-1}=1,\quad l\in T,
 \eqno{(1.2)}$$ 
where $T$ is a finite set of distinct real numbers
from the interval $(-0.5, \infty)$, and the sets $T$ and $S$ are
disjoint. This uniqueness result is crucial for the
arguments in \cite{CTo}.
For references on the inverse scattering with fixed-energy data 
see [1]-[9].
In this note a counterexample to the uniqueness claim from \cite{CT}
is constructed. This counterexample invalidates the
arguments in \cite{CT} and  \cite{CTo}.
The uniqueness of the solution of similar equations in \cite{CS}
(see equations (12.1.2) and (12.2.1) on pp.195-196 in \cite{CS})
and \cite{N} does not hold for some $r>0$, in
general, also.

\section{A counterexample} 

Take the single-element sets $S=\{0\}$ and $T=\{2\}$. In this case
equation (1.2) yields $\gamma_0=-6$,  
$g(s,t)=-6u_0(t)v_0(s):=-6g_0(s,t)$ for $t<s$ and $g_0=u_0(s)v_0(t)$
for $t>s$. Note that $(d^2+1)g_0=\delta (s-t)$, where $\delta (s-t)$
is the delta-function, and $d^2$ stands for the second derivative.
Therefore, applying the operator $d^2+1$ to equation (1.1),
one gets $(d^2+1)h=6s^{-2}h$. This equation has a nontrivial, regular
at zero, 
solution $cu_2(s)$, where $c\neq 0$ is a constant. Without loss
of generality we take $c=1$ in what follows.   

Note that $u_0(r)=\sin r, $ $v_0(r)=-\cos r$, 
$u_2(r)= (3r^{-2}-1)\sin r -3r^{-1} \cos r$, and the Wronskian
$u_0v'_0-u'_0v_0=1$.

Define the function 
$$ p(r):=v_0(r) u'_2(r)-v'_0(r) u_2(r).$$
One may check, using the explicit formulas for $v_0$ and $u_2$, that
$$p(r)=1-\frac {3+ 3\cos^2(r)}{r^2} +\frac {3\sin(2r)}{r^3} .$$
Using this explicit formula, one checks that 
$p(r)=-\frac {r^2} 5+ o(r^2)$ as $r\to 0,$
and that $p(r)>0$ as $r\to \infty$. In fact,
$p(r)\to 1$ as $r\to \infty$.
Since $p(r)$ is continuous on $(0, \infty)$,
one concludes that:

 {\bf There is a number $R>0$ such that $p(R)=0.$}

Let us now prove the following:

{\bf Claim: Equation (1.1) has a nontrivial solution $u_2(s)$
if $r=R$.}

 Proof of the claim:

One has:
$$-\int_0^r g(s,t)u_2(t)t^{-2}dt=\int_0^r g_0(s,t)6u_2(t)t^{-2}dt=
\int_0^r g_0(s,t)(d^2+1) u_2(t)dt:=J.$$
 Integrating by parts twice and taking into account that
$u_0$ and $u_2$ vanish at the origin, one gets:
$$J=u_2(s)[u_0(s)v'_0(s)-u'_0(s)v_0(s)]+u_0(s)[v_0(r)u'_2(r)-
v'_0(r)u_2(r)]=u_2(s) +p(r)u_0(s).$$
We have used above the formula for the Wronskian: $u_0v'_0-u'_0v_0=1$. 
 
It follows from the foregoing equation for $J$ that if, for some
$r>0,$ one
has $p(r)=0$, then $u_2(s)$ solves equation (1.1) with this $r>0$.

Since we have already proved that $p(R)=0$, the claim follows.
$\Box$

\end{document}